# Dynamics of electron-electron correlated to electron-phonon coupled phase progression in trilayer nickelate La$_4$Ni$_3$O$_{10}$

Sonia Deswal[1, †], Deepu Kumar[1], Dibyata Rout[2], Surjeet Singh[2] and Pradeep Kumar[1, *]

[1] *School of Physical Sciences, Indian Institute of Technology Mandi, Mandi-175005, India*

[2] *Department of Physics, Indian Institute of Science Education and Research Pune, Pune-411008, India*

**Abstract**

Trilayer nickelates are a rich class of materials exhibiting diverse correlated phenomena, including superconductivity, density wave transitions, non-Fermi liquid behavior along with an unusual metal-to-metal transition around T* ~ 150 K. Understanding the electronic correlations, lattice and charge dynamics are crucial to unreveal the origin of superconductivity and other instabilities in nickelates. Our in-depth Raman measurements shows that trilayer nickelate, La$_4$Ni$_3$O$_{10}$, shows transition from electron-phonon coupled phase to the electron-electron correlated one below charge density wave transition around T* with an estimated energy gap ($\Delta$) of ~ 18-20 meV. The transition around T* is also accompanied by the emergence of zone folded phonon modes reflecting the transition into the charge density wave phase. Phonon modes self-energy parameters show anomalous changes around T* attributed to the electron-electron correlations, and renormalization rate of the phonon modes is much slower in the charge-ordered phase compared to the phase above T*. The transition around T* are marked by the suppression of electron-phonon coupling parameter by ~ 70 %, a change of the quasiparticle dynamics from non-Fermi liquid to the Landau-Fermi liquid type behaviour estimated using the low frequency ($\omega \to 0$) Raman response.

[†]soniadeswal255@gmail.com
[*] pkumar@iitmandi.ac.in

## 1. Introduction

Transition metal oxides with perovskite-type structures have been the focus of intense investigation over the last few years due to the unique nature of outer *d*-electrons. Nickelates, cuprates and manganates are the most notable transition metal oxides that exhibit rich phenomena such as high-temperature superconductivity, pseudogap phase, charge and orbital ordering, and a plethora of other fascinating electronic and magnetic phases [1–4]. The coexistence of charge density wave (CDW) with the superconducting phase in high-$T_c$ cuprates remains an area of continued interest and focus [5–7]. Nickel based transition metals oxides, specifically Ruddlesden-Popper (RP) series $R_{n+1}Ni_nO_{3n+1}(n=1,2,3..........\infty)$ (R = rare earth), exhibit charge ordering, strongly coupled spin and lattice degree of freedom similar to cuprates, and both have striking electronic and structural similarities [8–10]. The physical properties of nickelates are reminiscent of high $T_c$ cuprate superconductors which makes them intriguing for studying the interplay between charge localization, lattice dynamics and their relationship with magnetism.

The metallic La$_4$Ni$_3$O$_{10}$ member of RP series (n=3) exhibit mixed valence states $Ni^{2+}/Ni^{3+}$ with an average valence of 2.67 [11,12]. One of the striking features of this material is metal-to-metal transition (MMT) around 140-150K [13–16], which is understood by the CDW instabilities associated with the common nesting vector of two hidden 1D Fermi surfaces [17]. La$_4$Ni$_3$O$_{10}$ possesses an Ni $d_{3z^2-r^2}$ band which shows a low-temperature energy gap of ~ 20 meV [18], whereas Ni $d_{x^2-y^2}$ orbitals show no pseudogap formation [19]. The evolution of $d_{3z^2-r^2}$ band gap from 120 K to 150 K is associated with MMT and play a crucial role in the charge and spin instabilities. It is also proposed that the incommensurate density waves with charge and spin characteristics on the Ni sites, similar to metallic density waves in chromium metal [20], drive the MMT transition [12]. Significant research efforts have been devoted to understand the precise mechanism underlying the interspersed charge and spin density that emerges at MMT and the complete understandig still lacks clarity. Another tantalizing observation in trilayer nickelates is the superconductivity under high pressure at ~ 20K [21–23]. As the underlying physics of cuprates is still not well understood and results into many surprises, therefore it is conjectured that nickelates may be pivotal in revealing the physics of cuprates as both have electronic and structural similarity.

Phonons and electronic degrees of freedom play a crucial role in controlling the physics of diverse systems including cuprates [24–30]. We note that the role of phonons and their coupling with the electronic degrees of freedom have not been explored in trilayer nickelates. Therefore, we have taken such a study to explore the phonon dynamics along with the electronic degrees of freedom using an in-depth temperature dependent Raman scattering measurements. Our measurements provide clear evidence of a MMT around 150 K, reflected in the appearance of a number of additional modes owing to zone folding and strong renormalization of the phonons self-energy parameters. Our measurements uncovered a transition from a strong electron-phonon coupled phase at high temperature to the electron-electron correlated state below T* (~ 150 K) marked by the transition from non-Fermi liquid (NFL) to the canonical Landau-Fermi liquid (FL) type behavior of the scattering rate of the quasiparticles estimated using the low-frequency Raman response.

**2. Results and discussion**

**2.1 Temperature dependence of the Phonon modes**

Here, we have used monoclinic $(P21/a)$ crystallographic symmetry for $La_4Ni_3O_{10}$. For this crystal symmetry, group-symmetry analysis yields 96 Raman active modes: $\Gamma_{Raman} = 48A_g + 48B_g$ [31] (see Table S1 and section S1 for more details on phonon modes and crystal structure, Supporting Information [32]). The Raman spectra of $La_4Ni_3O_{10}$ at few selected temperatures are shown in Fig. 1 (see Fig.S1 for mode labeling i.e. P1 to P33), which illustrates the disappearance of several phonon modes as temperature increases (see Supporting Information [32] Table S2 for modes frequencies). As the temperature increases and crosses T*, phonon modes exhibit two significant changes: first, quite a large number of phonon modes disappeared, see Fig. 2 (a); and second, nearly temperature independence of the phonon modes frequencies and full-width at half maxima (FWHM) in the CDW phase below T*, in sharp contrast to the renormalization above T*, see Fig. 2 (b) and (c). With increasing the temperature, numerous weak intensity phonon modes disappear in the temperature range of 120 to 160K [see Fig. 2 (a)]. These additional modes suggest their origin in the zone-folded phonon modes which are unique to the charge ordered phases [33,34]. Deep inside the CDW state (i.e. at T < T*), we observed well-distinct and sharp phonon modes. Frequency softening and linewidth broadening rate of the phonon modes increases drastically as the temperature increases above T*. We note that this is in sharp contrast to the simple anharmonic phonon-phonon model [35,36], which predicted a monotonous decreases/increase of the modes frequency/linewidth with increasing temperature. Our observation suggests that underlying physics in the CDW phase is controlled by the electron-electron correlations and above T* electron-phonon coupling start playing the role. In other words, as electron-phonon coupling is playing active role from room temperature till T*, so a strong renormalization of the phonon self-energy parameters, i.e. peak frequency and linewidth, is expected. Once below T* electron-electron correlations becomes active and phonon renormalization is expected to be minimum as electrons no longer effectively interact with the phonons. In particular, the intense phonon modes P8, P11, and P27 frequency softening rate is very small $\Delta\omega(\omega_{150K} - \omega_{6K}) \sim 2.5$ cm$^{-1}$ below $T*$. On the other hand, the rate of softening becomes almost double above T* i.e. $\Delta\omega(\omega_{330K} - \omega_{150K}) \sim 5$ cm$^{-1}$. Similarly, the linewidth broadening rate is small in the CDW phase i.e. $\Delta\Gamma(\Gamma_{150K} - \Gamma_{6K}) \sim 2$ cm$^{-1}$, however above T* it increases by a factor of three i.e. $\Delta\Gamma(\Gamma_{330K} - \Gamma_{150K}) \sim 6$ cm$^{-1}$. Above T*, the softening and broadening rate for the most intense mode P32 is very high as compared to the other modes i.e $\Delta\omega(\omega_{330K} - \omega_{150K}) \sim 15$ cm$^{-1}$ and $\Delta\Gamma(\Gamma_{330K} - \Gamma_{150K}) \sim 10$ cm$^{-1}$. These findings lead us to propose that the CDW phase exits below T* justifying the weak temperature dependency observed below T* as compared to that above T*. Figure 2 (d) shows temperature evolution of the intensity for the modes P8, P11, P27, and P32 normalized to their intensity at 6 K. The intensity decreases slowly as the temperature increases till T* and above T* it changes rapidly (see Fig. S2 in Supporting Information [32] ).

The trend of very weak softening and broadening rate below T* is attributed to the emergence of electron-electron correlation in the CDW phase, which is also believed to be the driving force behind the formation of CDW in La$_4$Ni$_3$O$_{10}$. In a simplistic picture, to quantify the phonons anomalies in the CDW phase and to understand the effect of cubic anharmonicity (decay of an optical phonon into two acoustic phonons of equal frequency) on the phonon modes frequencies and FWHM in the stable metallic phase (150 – 330K) is assessed by fitting with the anharmonic phonon-phonon interaction model given as [35,36]:

$$\omega(T) = \omega_0 + A\left(1 + \frac{2}{e^x - 1}\right), \text{ and } \Gamma(T) = \Gamma_0 + B\left(1 + \frac{2}{e^x - 1}\right)$$

where $\omega_0$, and $\Gamma_0$ are the mode frequency and FWHM at absolute zero temperature, respectively and $x = \frac{\hbar\omega_0}{2\kappa_B T}$, and constant parameters A and B (see Supporting Information [32] Table S3 for fitting parameters). Temperature dependence of the frequency and FWHM of the prominent phonon modes as shown in Fig 3 (a) and (b), for other modes see Supporting Information [32] Fig. S3, fits well from 330 - 150K (solid red curve). Below T* we extrapolated this anharmonic curve, see dotted lines, till the lowest recorded temperature. It's apparent that below ~ 150K, phonon modes exhibit abnormal characteristics and deviate from the predictions of a simple anharmonic model. The Linewidth of these modes shows a sharp decrease below T*. This anomalous behavior of the linewidths suggests that electron-electron correlations are at play in the CDW phase. As a result, electron-phonon coupling is minimal and which may give rise to longer phonon lifetime ($\tau$) and hence this decrease in the phonon linewidth ($\propto 1/\tau$) below T*.

Renormalization of the phonon frequency below T* may also be due to Fermi surface reconstruction as a result of gap opening provided phonon couples to that part of the Fermi surface. In a simplistic picture, phonon mode frequency is expected to increase with decreasing temperature owing to reduced bond length ($r$) or positive lattice expansion, as $\omega \propto 1/(r)^{3/2}$. However, we do observed phonon mode softening for some of the modes below T* and this softening may results due to negative lattice expansion. Indeed a negative thermal expansion has been reported along *b*-axis for this system below T* [14,15], though the effect of lattice expansion ($\Delta b/b \sim 0.02\%$) on phonon softening is expected to be small, $\Delta \omega_{max} \sim 0.2$ cm$^{-1}$. Therefore, the effect of Fermi surface renormalization is expected to dominate. A detailed quantitative analysis of the Fermi surface renormalization on the phonon modes softening demands a thorough theoretical study for these systems. We note that strength of the electron-phonon coupling can be estimated using renormalization of the phonon linewidths for the case of metallic systems [37]. For the n$^{th}$ phonon having FWHM - $\Gamma_n$ and peak frequency - $\omega_n$, they are related as: $\frac{\Gamma_n}{\omega_n^2} = \frac{2\pi}{g_n}\lambda_n N_F$, where $g_n$ is the degeneracy of the n$^{th}$ phonon mode which is one here and $\lambda_n$ is the electron-phonon coupling. Figure 3 (c) shows the temperature evolution of $\frac{\Gamma_n}{\omega_n^2}$ for the modes P11, P14, P27, and P32. We see clear variation in the vicinity of T* showing that $\lambda_n$ decreases by 50 to 70% from room temperature to the low temperature. Taking the estimated value of $N_F$ = 0.97/1.03 states/eV [21] at room/low temperature, and using $\frac{\Gamma_n}{\omega_n^2}$ values at room/low temperature, we estimated the electron-phonon coupling constant for these modes, see Table S4. The estimated electron-phonon coupling constant from our measurements is of similar magnitude as reported earlier both theoretically and experimentally via transport measurements [14,38]. We note the magnitude of the electron-phonon coupling decreases drastically below T*, suggesting electrons becomes more coherent with decreasing temperature and resulting into more increased electron-electron correlations, signalling a tendency of gravitating towards the Landau FL regime as also evidenced by the scattering rate estimation described below in section 2.2.

Further, we discuss the low-frequency phonon modes (P1 - P6) shows asymmetric line shape, prominent at high temperatures, and the origin of this asymmetry may be captured by the Fano model [39]. Our observation of asymmetric line shapes for the phonon modes is attributed to the quantum interference between the discrete phonon excitation and the underlying electronic continuum. As a result of mode splitting and the emergence of new modes at low temperatures (see Supporting Information [32] Fig. S4), we are unable to quantify independent Fano asymmetry characteristics for any particular mode. However, the low-frequency bundle of modes (P1-P6) exhibits clear asymmetry at high temperature, which is evident by the well-fitted (see blue line as a fit in the inset of Fig. 4. (a)) with the Breit-Wigner-Fano (BWF) line shape [39,40] $I_{BWF}(\omega) = \frac{I_0 [1+(\omega-\omega_c)/q\Gamma]^2}{1+[(\omega-\omega_c)/\Gamma]^2}$, where $\omega, \omega_c, \Gamma,$ and $I_0$ are the Raman shift, the spectral peak center, FWHM, and intensity, respectively; $q$ is the asymmetry parameter [39]. Microscopically, $q$ is inversely proportional to the interaction parameter i.e. $|q| \propto 1/|V_E|$; where $V_E (= \langle \psi_E | H | \phi \rangle)$ is a transition which provides the information about the interaction between the discrete state and the continuum and $2\pi |V_E|^2$ represents the band width of the unperturbed continuum states. Therefore, when $1/|q| (\propto |V_E|) \to \infty$; coupling is quantified as stronger while in the limit $1/|q| \to 0$; coupling is weak and this results into the Lorentzian line shape. $q$, is linked to the coupling strength and is intimately connected with the electronic polarizability. Therefore, any change in the underlying electronic correlations is also expected to be reflected in $q$. Figure 4 (a) shows the temperature dependence of the coupling strength ($1/|q|$) in the temperature range of 90-330K, we limited our fitting with Fano function till ~ 90K as the modes splitting becomes prominent below ~ 90K and start to gain Lorentzian shape. We note that $1/|q|$ is very small at low temperature below T* reflecting weak electron-phonon coupling. However, it increases linearly above T* till room temperature suggesting strong electron-phonon coupling in this regime. We note that due to electron-electron correlation at low temperature there is a possible splitting in the bandwidth with in the Hubbard model and results in to the effective decrease in the bandwidth. This splitting may lead to reduction in the bandwidth of the continuum given as $2\pi |V_E|^2$ and as a result be responsible for the reduction in the $1/|q|$ as it is directly proportional to the interaction strength between electrons and phonons. Therefore, from our observation the stagnation of the interaction parameter below T* hints emergence of the electron-electron correlated phase from the high temperature strong electron-phonon coupled regime; and one should expect phonon driven transport above T* and electron-electron correlation driven at low temperature.

**2.2 Spectral weight redistribution and charge carrier dynamics**

The signature of emergent charge dynamics may also be gauged via renormalization of the spectral weight below the CDW ordering temperature. For metallic systems the dynamics of charge and spin fluctuations may be probed by Raman scattering spectral weight redistributions [26,41,42]. Fig. 4 (b) illustrates Raman response $\chi''(\omega,T) \propto I(\omega,T)/1+n(\omega,T)$ the continuum extending up to 750 cm$^{-1}$, which shows a notable shift in spectral continuum from low to high frequencies as we increase the temperature. The emergence of a CDW energy gap ($\Delta$) in the electronic excitation spectrum is discernible through progressive redistribution of the Raman spectral weight [43–45]. This proposed approach allows us to estimate the CDW energy gap value, which is close to the intersection points $2\Delta \sim 36\text{-}40$ meV in $\chi''(\omega,T)$, indicated by a black arrow in Fig.4 (b). A quantitative analysis of the Raman's response can be done by integrated intensity $I_{\chi''(T)}$ over the full range [8 ($\omega_l$) to 1350 ($\omega_h$) cm$^{-1}$]. $I_{\chi''(T)}$ remains nearly unchanged above T*, while below T*, continuous reduction in $I_{\chi''(T)}$ is observed with lowering the temperature [shown in Fig. 4(c)] suggesting opening of the gap. The estimated CDW energy gap derived from the spectral weight redistribution is $\Delta \sim 18-20\ meV$, (see Fig.4 (b)). Similar values for the CDW gap $\sim 20\ meV$ have been reported previously using nuclear spin relaxation, and ARPES measurements [18,46].

The progression in the strength of electron-electron correlation can also be gauge from the low frequency ($\omega \to 0$) Raman response. Quantitatively, dynamics of underlying quasiparticle may be extracted from the low frequency slope of the Raman response. The low-frequency response is given as [47–50]:

$$\chi''(\omega \to 0) = \omega N_F \left\langle \frac{\gamma^2(k) \int d\omega (-\partial f^0/\partial \omega) Z_k^2(\omega,T)}{2 \sum_k''(\omega,T)} \right\rangle$$

Where $N_F$ is the density of electronic levels at the Fermi level, $\gamma(k)$ is the scattering amplitude. $\sum_k^{''}$ is imaginary part of the self-energy related to the quasiparticle lifetime ($\tau_k(\omega,T)$) as $\hbar/2\sum_k^{''}(\omega,T) = \tau_k(\omega,T)$ and $Z_k(\omega,T) = [1-\partial\sum_k^{'}(\omega,T)/\partial\omega]^{-1}$ is the quasiparticle residue. $f^0$ is the equilibrium Fermi distribution function, and $\langle \rangle$ represents an average over the Fermi surface. Therefore, the low frequency Raman slope is proportional to the lifetime ($\tau_0(T)$) or inversely proportional to the scattering rate, $\Gamma_0(T)$, of the low energy conduction electrons i.e. $\left.\frac{\partial\chi''(\omega,T)}{\partial\omega}\right|_{\omega=0} = \frac{1}{\Gamma_0(T)} \propto \tau_0(T)/\hbar$, and it effectively measures the scattering rate of the conduction electrons in a metal also referred as Raman resistivity. It is anticipated that with increasing correlation at low temperature, as the system under study goes from metal to correlated metal phase transition, one expects a canonical Landau FL type behaviour of the putative quasiparticles i.e. scattering rate $\propto T^2$. For higher temperature, larger than the coherence scale, quasiparticles become short-lived and the Landau Fermi liquid description no longer hold. Figure 5(a-b) shows the inverse of slope i.e. $\left[\frac{\partial\chi''(\omega)}{\partial\omega}\right]^{-1}$ as a function of temperature. We estimated the slope using two different ways, first by linear fitting in the low frequency region i.e. from 8 to 25 cm$^{-1}$ (see Fig. 5(b)). Second by considering the large frequency range, till ~ 50 cm$^{-1}$, and fitting the Raman response using a polynomial i.e. $\chi'' \propto m_1\omega + m_2\omega^2$. We extracted the slope using the procedures described above [see Fig. S5 for the fitting and Fig. 5(a) for the extracted slope $m_1$]. We note that the qualitative temperature evolution of the slope extracted via different procedure is similar. At high temperature where electro-phonon coupling dominates, i.e. electron correlation energy is zero or negligible, the inverse of slope is linear in temperature suggesting a NFL behaviour.

With decreasing temperature below T*, there is a finite electron-electron correlated energy (U) contribution and the correlated metal displays an inverse slope $\propto T^2$ reflecting a canonical Landau FL type behaviour. We note that the inverse slope shows a change below ~ 20K suggesting the beginning of an insulating behaviour as the temperature decreases. Interestingly, in the transport measurements below this temperature resistivity upturn has been reported, which was attributed to the pure electron-electron based transport and possibly a metal to insulator transition [14,15]. Our results established the evolution from an electron-electron correlated regime at low temperature to the electron-phonon coupled phase at high temperature for this system. The increase in the electron-electron correlation at low temperature is also supported by the earlier experimental reports of enhanced effective mass which increases below T* [14,15].

## 3. Conclusion

In conclusions, our studies reveals strong electronic correlations in the CDW phase below T* marked by different parameters qualitatively estimated here. The quasiparticle scattering changes from non-Fermi liquid to the Landau-Fermi liquid type behaviour below T*. The spectral redistribution reveals a CDW energy gap $\Delta \sim 18-20\, meV$. Our results suggest electron-electron correlations in the CDW phase below T* and strong electron-phonon coupling in the high temperature phase, reflected in the renormalized phonon modes and intriguing temperature evolution of the Raman response. Our results provide key information for understanding the nature of the transition at T* in trilayer nickelate $La_4Ni_3O_{10}$. We anticipate that our work may help in uncovering the underlying physics which may be even useful for cuprates.

**Materials and Methods:**

The polycrystalline sample was synthesized and characterized as described in Ref. [15] Unpolarized temperature-dependent Raman measurements were performed in backscattering geometry using a Horriba HR Evolution spectrometer equipped with Peltier-cooled charged coupled device detector to collect the scattered light. The sample was excited with a solid-state laser of 532 nm wavelength, to prevent any impacts of local heating the laser power was kept very low, < 1 mW. A 50× long working distance objective was utilized to focus the laser on the sample surface as well as to collect the scattered light. For the higher spectral resolution, we used a higher groove density grating (1800 lines/mm) which disperses the Raman signals onto the detector. A closed cycle He cryostat (Montana Instruments) was used for variation of temperature in the range from 6 K to 330 K while maintaining temperature stability of ~ ± 0.1 K, under the vacuum of ~ 90 $\mu Torr$

**Acknowledgement:** PK acknowledge financial support from Science and Engineering Research Board (Science and Engineering Research Board - Project no. CRG/2023/002069) and IIT Mandi for the experimental facilities.

**References:**

[1]     H. J. Zhao, W. Ren, Y. Yang, J. Íñiguez, X. M. Chen and L. Bellaiche, Nat. Commun. **5**, 4021 (2014).

[2]     M. Hepting, M. Minola, A. Frano, G. Cristiani, G. Logvenov, E. Schierle, M. Wu, M. Bluschke, E. Weschke, H. Habermeier, E. Benckiser, M. Le Tacon and B. Keimer, Phys. Rev. B **113**, 227206 (2014).

[3]     P. Sippel, S. Krohns, E. Thoms, E. Ruff, S. Riegg, H. Kirchhain, F. Schrettle, A. Reller and A. Loidl, The European Phy. Journal B **85**, 235 (2012).

[4] K.-W. Lee and W. E. Pickett, Phys. Rev. B **70,** 165109 (2004).

[5] M. Vojta, Adv. Phys. **58**, 699 (2009).

[6] J. Chang, E. Blackburn, A.T. Holmes, N.B. Christensen, J. Larsen, J. Mesot, Ruixing, D.A. Bonn, W.N. Hardy, A. Watenphul, M.v. Zimmermann, E.M. Forgan and S.M. Hayden Nat. Phys. **8**, 871 (2012).

[7] Y. Yu and S. A. Kivelson, Phys. Rev. B **99**, 144513 (2019).

[8] V.I. Anisimov, D. Bukhvalov and T. M. Rice, Phys. Rev. B **59**, 12 (1999).

[9] A. S. Botana and M. R. Norman, Phys. Rev. X **10**, 011024 (2020).

[10] M. C. Jung, J. Kapeghian, C. Hanson, B. Pamuk, and A. S. Botana, Phys. Rev. B **105**, 085150 (2022).

[11] M. D. Carvalho, M. M. Cruz, A. Wattiaux, J. M. Bassat, F. M. A. Costa, and M. Godinho, Appl. Phys. **88**, 1 (2000).

[12] J. Zhang, D. Phelan, A. S. Botana, Y. Chen, H. Zheng, M. Krogstad, S. G. Wang, Y. Qiu, J.A. Rodriguez-Rivera, R. Osborn and S. Rosenkranz, M. R. Norman, J. F. Mitchell, Nat. Commun. **11**, 6003 (2020).

[13] M. Greenblatt, Solid State and Mater. Sci. **2**, 174 (1997).

[14] S. Kumar, Ø. Fjellvåg, A. O. Sjåstad and H. Fjellvåg, Journal of Magnetism and Magnetic Mat. **496**, 165915 (2020).

[15] D. Rout, S. R. Mudi, M. Hoffmann and S. Spachmann, R. Klingeler, and S. Singh, Phys. Rev. B **102**, 195144 (2020).

[16] J. Zhang, H. Zheng, Y. Chen, Y. Ren, M. Yonemura, A. Huq and J. F. Mitchell, Phys. Rev. Mater. **4,** 083402, (2020).

[17] D. K. Seo, W. Liang, M. H. Whangbo, Z. Zhang, and M. Greenblatt, Inorg. Chem**. 35**, 6396 (1996)

[18] H. Li, X. Zhou, T. Nummy, J. Zhang, V. Pardo, W. E. Pickett, J. F. Mitchell and D. S.

Dessau, Nat. Commun. **8**, 704 (2017).

[19]     M. Imada, A. Fujimori, and Y. Tokura, Rev. Mod. Phys. **70**, 1039 (1998).

[20]     Eric Fawcett. Phys. Rev. B **60,** 209 (1988).

[21]     Qing Li, Y.-Jie Zhang, Z.-Ning Xiang, Y. Zhang, X. Zhu and Hai-Hu Wen, (2023) https://arxiv.org/abs/2311.05453 .

[22]     J.X. Wang, Z. Ouyang, R.-Q. He, and Z.-Y. Lu, Phys. Rev. B. **109**, 165140, (2024).

[23]     Y. Zhu, D. Peng, E. Zhang, *et al.*, Nature **631**, 531 (2024).

[24]     P. Kumar, A.Kumar, S. Saha, D.V.S Muthua, J.Prakash, S. Patnaik, U.V. Waghmare, A.K. Ganguli, and A.K. Sood, Solid State Comm. **150,** 557 (2010).

[25]     B. Singh, M. Vogl, S. Wurmehl, S. Aswartham, B. Büchner, and P. Kumar, Phys. Rev. Research **2**, 013040 (2020).

[26]     B.Loret, N. Auvray, Y. Gallais, M. Cazayous, A. Forget, D. Colson, M. H. Julien, I. Paul, M. Civelli, and A. Sacuto, Nat. Phys. **15**, 771 (2019).

[27]     P. Kumar, D.V.S. Muthu, L. Harnagea, S. Wurmehl, B. Buchner, and A.K. Sood, J. Phys.: Condens. Matter **26**, 305403 (2014).

[28]     D. Kumar, B. Singh, R. Kumar, M. Kumar, and P. Kumar, J. Phys.: Condens. Matter **32**, 415702 (2020).

[29]     T. Strohm and M. Cardona, Phys. Rev. B - Condens. Matter Mater. Phys. **55**, 12725 (1997).

[30]     P. Kumar, A. Bera, D.V. S Muthu, P.M. Shirage, A. Iyo, and A.K. Sood, Appl, Phys. Lett. **100**, 222602 (2012).

[31]     D.L.Rousseau, R. P. Bauman, and S.P.S. Porto, J. Raman spectrosc., **10**,253 (1981).

[32]     See the Supporting Information at [URL will be inserted by publisher] for details on the experimental techniques, along with additional figures and tables for better understanding.

[33] J. Joshi, H. M. Hill, S. Chowdhury, C. D. Malliakas, F. Tavazza, U. Chatterjee, A. R. Hight Walker, and P. M. Vora, Phys. Rev. B **99**, 245144 (2019).

[34] G. Liu, X. Ma, K. He, Q. Li, H. Tan, Y. Liu, J.Xu, W. Tang, K. Watanabe, T. Taniguchi, L. Gao, Y. Dai, H. Wen, B. Yan, and X. Xi, Nat. Commun. **13**, 3461 (2022).

[35] M. Balkanski, R. F. Wallis, and E. Haro, Phys. Rev. B **28**, (1983).

[36] P.G. Klemens, Phys. Rev. B. **148**, 845, (1966).

[37] P. B. Allen, Solid State Commun. **14**, 937 (1974).

[38] J. Zhan, Y. Gu, X. Wu, and J. Hu, (2024) https://arxiv.org/abs/2404.03638.

[39] U. Fano, Phys. Rev. **124**, 1866 (1961).

[40] E. H. Hasdeo, A. R. T. Nugraha, M. S. Dresselhaus, and R. Saito, Phys. Rev. B **90**, 245140, (2014).

[41] K. Sen, D. Fuchs, R. Heid, K. Kleindienst, K. Wolff, J. Schmalian, and M. Le Tacon, Nat. Commun. **11**, 4270 (2020).

[42] Z. Liu, M. Huo, J. Li, Qing Li, Y. Liu, Y. Dai, X. Zhou, J. Hao, Yi Lu, M. Wang, and H. Wen (2024) https://arxiv.org/abs/2307.02950.

[43] H. M. Eiter, M. Lavagnini, R. Hackl, E. A. Nowadnick, A. F. Kemper, T. P. Devereaux, J. H. Chu, J. G. Analytis, I. R. Fisher, and L. Degiorgi, Proc. Natl. Acad. Sci. 110, 64 (2013).

[44] G. He, L.Peis, E. Cuddy, Z.Zhao, D.Li, R. Stumberger, B. Mortiz, H.T. Yang, H.J.Gao, T.P. Devereaux, and R.Hackl, Nat. Commun. **15**, 1895 (2024).

[45] N. Auvray B. Loret, S. Benhabib, M. Cazayous, R.D. Zhong, J. Schneeloch, G.D Gu, A. Forget, D. Colson, I. Paul, A. Sacuto and Y. Gallais, Nat. Commun. **10**, 5209 (2019).

[46] M. Kakoi, T. Oi, Y. Ohshita, M. Yashima, K. Kuroki, Y. Adachi, N. Hatada, T. Uda, and H. Mukuda, (2023) http://arxiv.org/abs/2312.11844v1.

[47] T. P. Devereaux and A. P. Kampf, Phys. Rev. B 5**9**, 6411 (1999).

[48]     F. Venturini, M. Opel, T. P. Devereaux, J. K. Freericks, I. Tüttő, B. Revaz, E. Walker, H. Berger, L. Forró, and R. Hackl, Phys. Rev. Lett. **89**, 107003 (2002).

[49]     T. P. Devereaux and R. Hackl, Rev. Mod. Phys. **79**, 175 (2007).

[50]     M. Opel, R. Nemetschek, C. Hoffmann, R. Philipp, P. Müller, R. Hackl, I. Tüttő, A. Erb, B. Revaz, E. Walker, H. Berger and L. Forro, Phys. Rev. B **61**, 9752 (2000).

**Figures:**

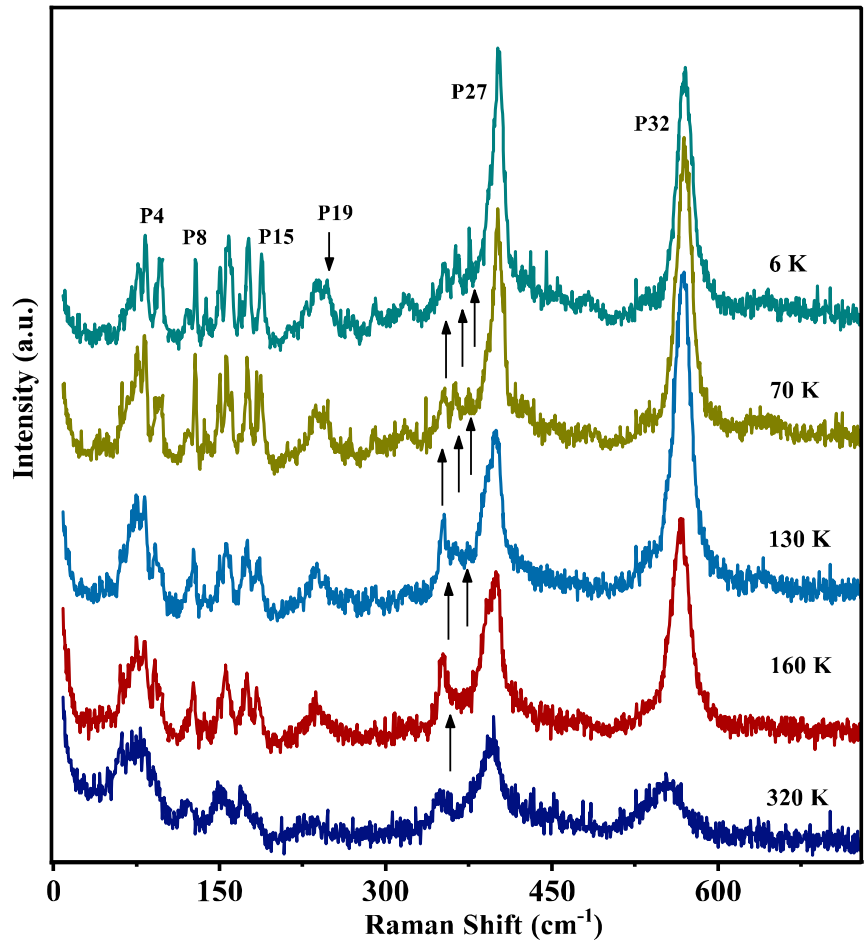

**Figure 1:** Raw Raman spectra of $La_4Ni_3O_{10}$ excited by 532 nm laser at different temperatures. The arrow indicates the disappearance of modes as temperature increases.

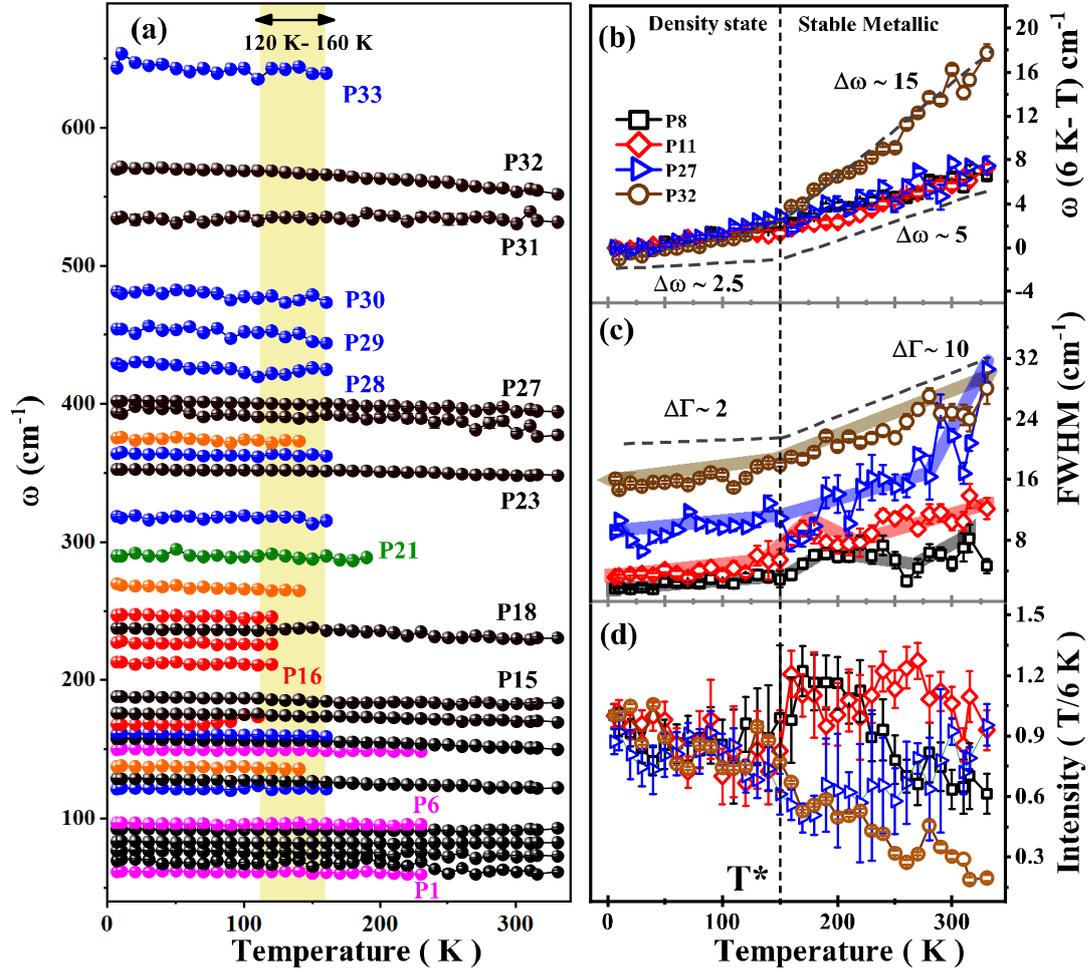

**Figure 2:** (a) Temperature dependence of the mode frequencies. Yellow-shaded area indicate the temperature range (120-160 K) where approximately half of the recorded phonon modes disappeared. (b-d) Temperature evolution of the frequency difference with respect to the frequency at 6 K i.e. $\omega_{6K} - \omega_T$, FWHM, and normalized integrated intensity with respect to the 6 K value i.e. $Inten._{Temp.} / Inten._{6K}$, respectively, for the prominent modes.

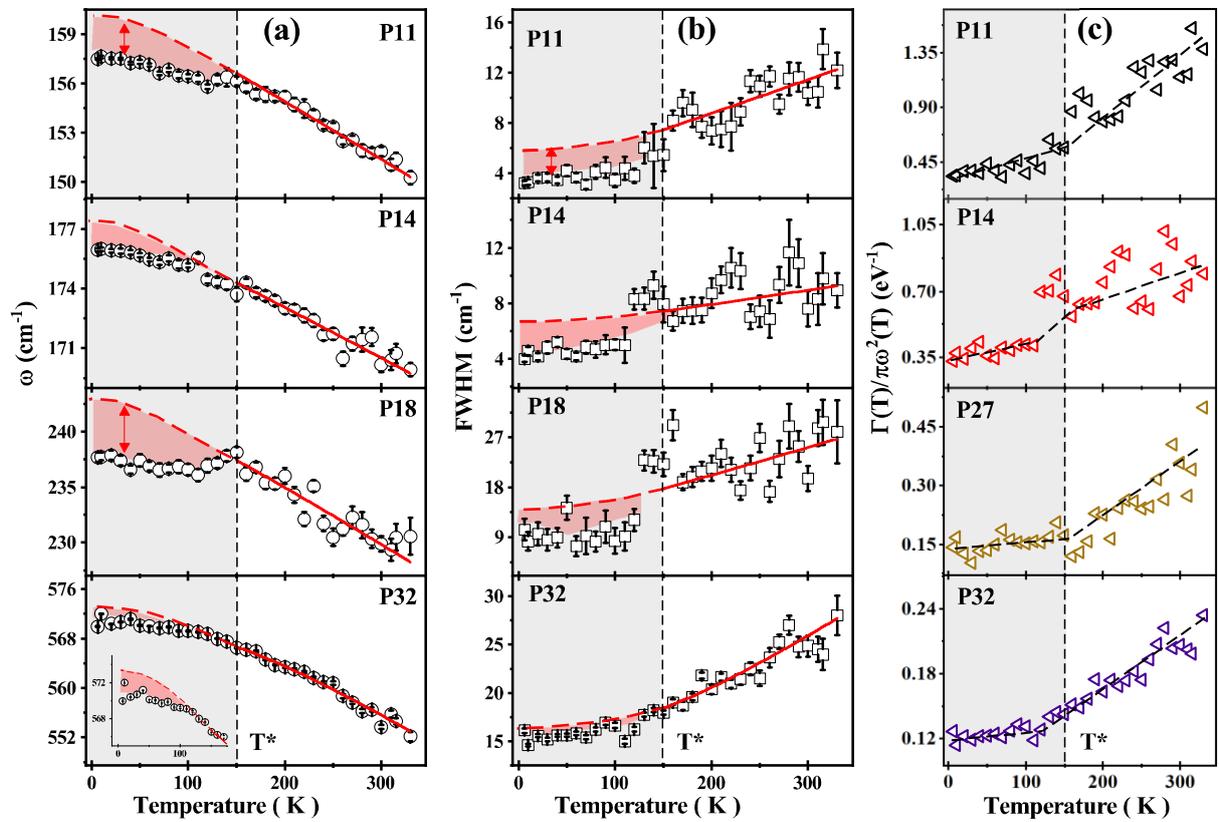

**Figure 3:** (a), and (b) Temperature variation of frequencies $(\omega)$ and FWHM for some of the prominent phonon modes, respectively. Solid red lines are fit to the anharmonic model as described in the text, and broken red lines are the extrapolated fitted curves. (c) $\Gamma(T)/\pi\omega^2(T)$ as a function of temperature for some of the prominent phonon modes, black dash lines are drawn as a guide to the eye. Shaded regions demonstrate the density wave order state.

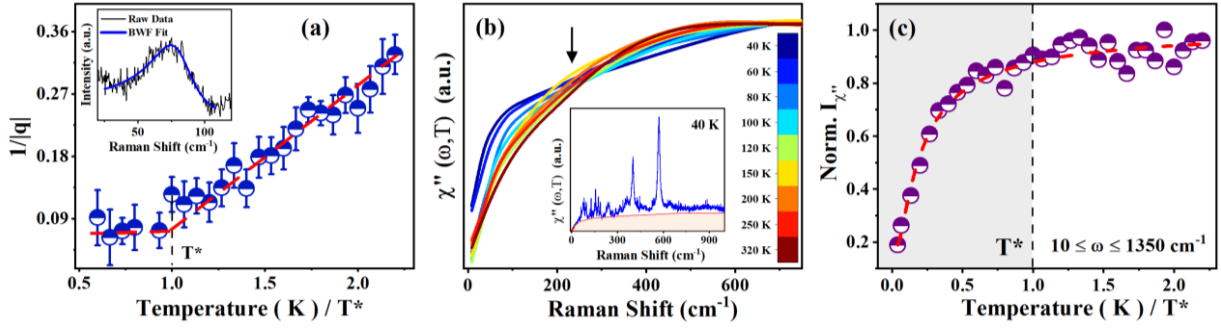

**Figure 4:** (a) $1/|q|$ as a function of normalized temperature T/ T*, dashed red line is guide to the eye and the dashed black line marks $T* \sim 150$ K. Inset shows the fit of the phonons cluster using BWF line shape. (b) The spectral continuum at different temperature, the one extracted at 40 K is shown by the red line in the inset. Inset shows the 40 K Raman response spectra; the shaded region demarcated by solid red line indicate the continuum. (c) Integrated intensity $I_{\chi"}$ of the contiuum in the frequency range, 8-1350 $cm^{-1}$. Dashed red line is guide to the eye.

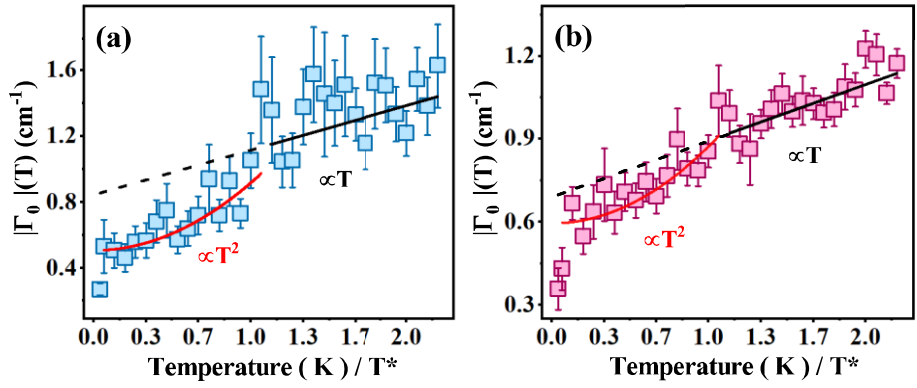

**Figure 5:** Raman relaxation rates as a function of temperature obtained from (a) polynomial and (b) linear initial slope. The solid red and black lines are quadratic and linear fits, respectively. Dashed black lines are extrapolations of the linear fitted curve.

**Suppporting Information:**

# Dynamics of electron-electron correlated to electron-phonon coupled phase progression in trilayer nickelate La$_4$Ni$_3$O$_{10}$

Sonia Deswal[1, †], Deepu Kumar[1], Dibyata Rout[2], Surjeet Singh[2] and Pradeep Kumar[1, *]

[1] *School of Physical Sciences, Indian Institute of Technology Mandi, Mandi-175005, India*
[2] *Department of Physics, Indian Institute of Science Education and Research Pune, Pune-411008, India*

[†]Soniadeswal255@gmail.com
[*] pkumar@iitmandi.ac.in

**S1. Phonon modes in La$_4$Ni$_3$O$_{10}$**

Several crystal structure studies, on La$_4$Ni$_3$O$_{10}$, performed using high-resolution temperature-dependent *X*-ray diffraction suggest two phases with four formula units (Z = 4). One is with an orthorhombic structure having space group *Bmab* and second, the dominating one, is the monoclinic structure with a space group $P21/a$ [1–6]. However, the crystal structure of La$_4$Ni$_3$O$_{10}$ remains up for debate, despite the previous studies. We note that the deviation of these different symmetries from each other is very small. Here, we have considered monoclinic $(P21/a)$ crystallographic symmetry based on the Rietveld profile refinement of high-resolution synchrotron based temperature-dependent X-ray diffraction data on the same set of high quality samples as used in our measurements [4]. For this crystal symmetry, group-symmetry analysis yields 96 Raman active modes: $\Gamma_{Raman} = 48A_g + 48B_g$ [7] (see Table S1 for more details).

**References:**

[1] D. Puggioni and J. M. Rondinelli, Phys. Rev. B **97**, 115116 (2018).

[2] C. D. Ling and D. N. Argyriou, Journal of Solid-State Chemistry **152**, 525 (2000).

[3] A. Olafsen, H. Fjellva and B. C. Hauback, Journal of Solid-State Chemistry **151,** 46 (2000).

[4] D. Rout, S. R. Mudi, M. Hoffmann and S. Spachmann, R. Klingeler, and S. Singh, Phys. Rev. B **102**, 195144 (2020).

[5] S. Kumar, Ø. Fjellvåg, A. O. Sjåstad and H. Fjellvåg, Journal of Magnetism and Magnetic Mat. **496**, 165915 (2020).

[6] J. Zhang, D. Phelan, A. S. Botana, Y. Chen, H. Zheng, M. Krogstad, S. G. Wang, Y. Qiu, J.A. Rodriguez-Rivera, R. Osborn and S. Rosenkranz, M. R. Norman, J. F. Mitchell, Nat. Commun. **11**, 6003 (2020).

[7] D.L.Rousseau, R. P. Bauman, and S.P.S. Porto, J. Raman spectrosc., **10**,253 (1981).

**Figures:**

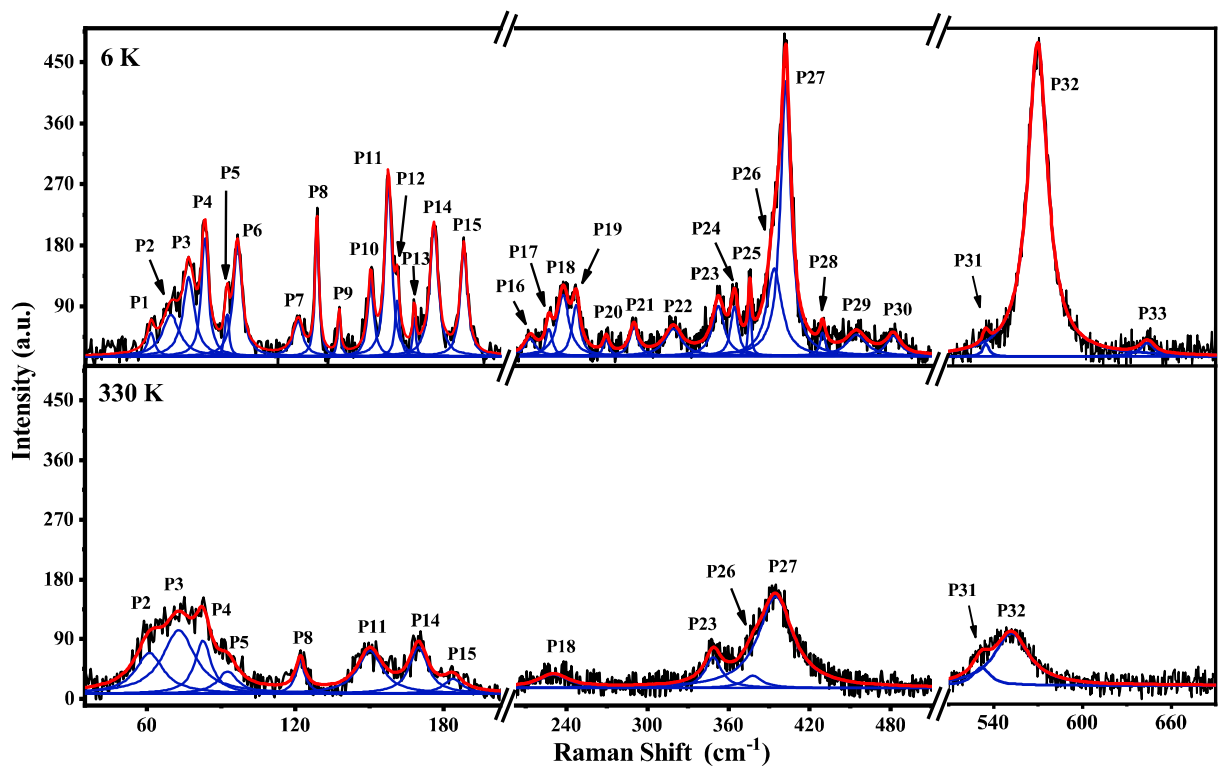

**Figure S1:** Raman spectra of $La_4Ni_3O_{10}$ in the spectral range of 35-700 cm$^{-1}$ collected at 6 K and 330 K. The solid red thick line is a total sum of Lorentzian fit of the experimental data (black colour) and solid thin blue lines correspond to the individual fit of the phonon modes. The observed modes are labeled as P1-P33.

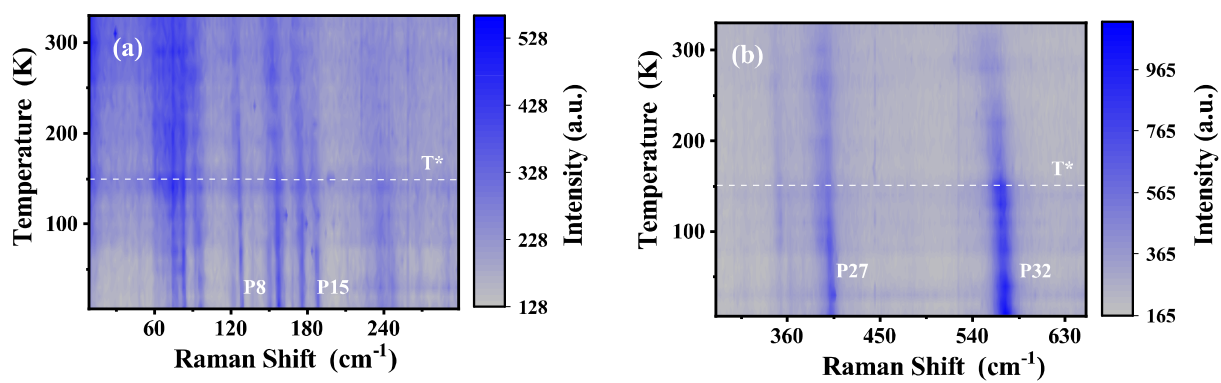

**Figure S2:** The color map in (a) and (b) illustrates that the strip of signals below T* begins to amalgamate as the temperature rises toward room temperature. The intensity abruptly drops for all phonon modes, and some of the modes completely disappear upon increasing the temperature above T*.

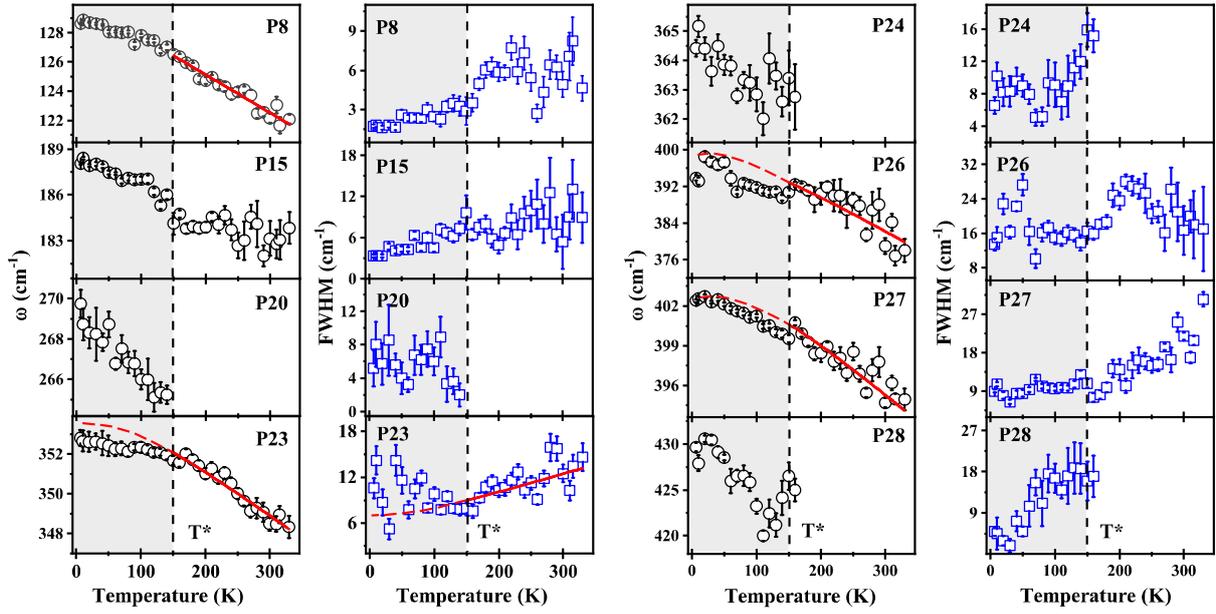

**Figure S3:** Temperature dependence of the frequency $(\omega)$ and full width at half maxima (FWHM) for the phonon modes. Shaded regions demonstrate the density wave order state. Error bars are the standard deviation obtained from the Lorentzian fits to the phonon peaks. The dashed black line at T* (~ 150 K) marks the CDW transition temperature. Solid red lines are fit to the anharmonic model as described in the text, and broken red lines are the extrapolated fitted curves.

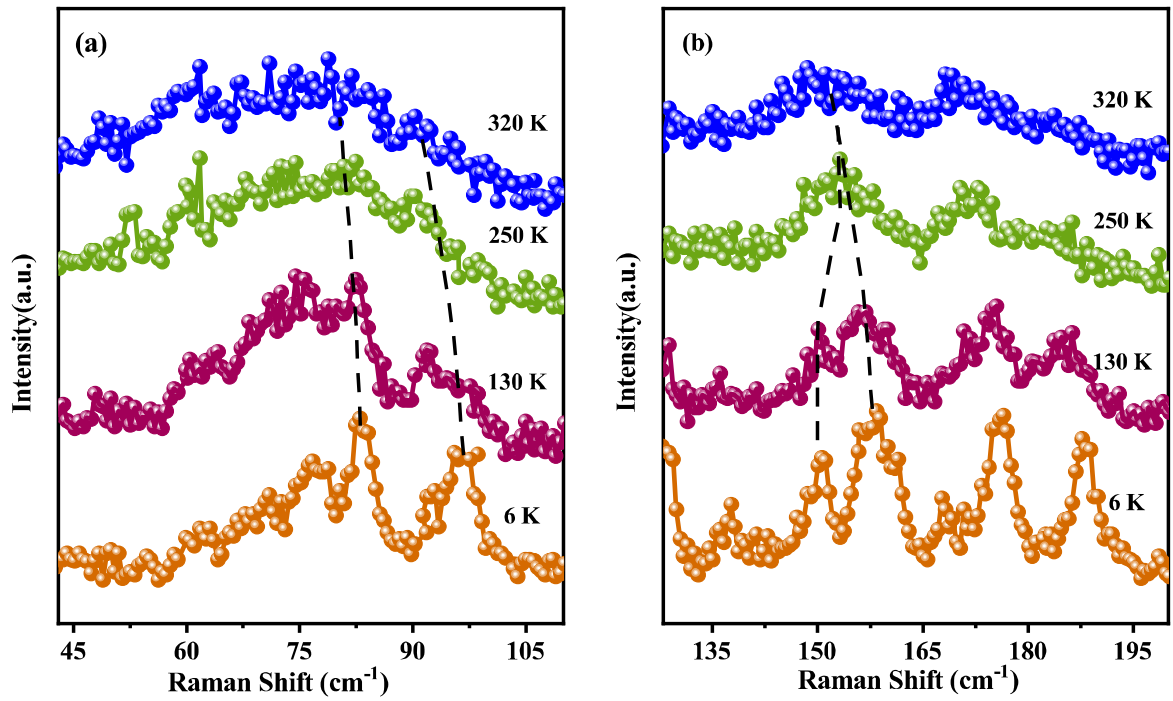

**Figure S4:** (a), (b) Raman spectra at different temperatures showing splitting of the phonon modes in the spectral range of 43-110 cm$^{-1}$ and 130-200 cm$^{-1}$, respectively.

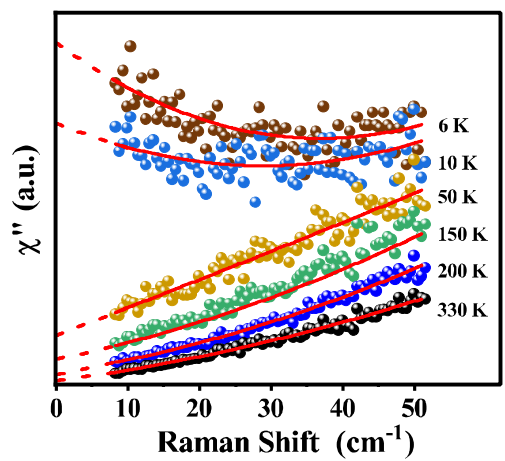

**Figure S5:** Low-frequency stokes-Raman response at different temperatures. Solid lines are the polynomial fits to the data: dashed lines indicate extrapolations to $\omega \rightarrow 0$.

**Tables:**

**Table S1:** Atoms and the corresponding Wyckoff positions in the unit cell and irreducible representations of the phonon modes at $\Gamma$ point for the monoclinic (space group $P2_1/a$) trilayer nickelate $La_4Ni_3O_{10}$. Irreducible representation $\Gamma_{Total}, \Gamma_{Raman}, \Gamma_{IR}$ and $\Gamma_{Acoustic}$ corresponds to total, Raman, Infrared, and acoustic phonon modes representation at the $\Gamma$ point, respectively. The second digit in the bracket in the first column represents the number of that particular atom in the unit cell.

| Monoclinic $P2_1/a$ (14) | | |
|---|---|---|
| **Atom** | **Wyckoff site** | **Point mode Decomposition** |
| **La[1,3]** | 4e | $3A_g + 3A_u + 3B_g + 3B_u$ |
| **Ni(1,2)** | 2a | $3A_u + 3B_u$ |
| **Ni(3)** | 2b | $3A_u + 3B_u$ |
| **Ni(4)** | 4e | $3A_g + 3A_u + 3B_g + 3B_u$ |
| **O[1,10]** | 4e | $3A_g + 3A_u + 3B_g + 3B_u$ |

$\Gamma_{Total} = 48A_g + 54A_u + 48B_g + 54B_u, \Gamma_{Raman} = 48A_g + 48B_g, \Gamma_{IR} = 53A_u + 52B_u, \Gamma_{Acoustical} = A_u + 2B_u$

**Table S2:** List of the experimentally observed modes with their frequencies at 6 K and 330 K. Unit in cm$^{-1}$.

| Mode Assignment | $\omega$ 6 K | $\omega$ 330 K | Mode Assignment | $\omega$ 6 K | $\omega$ 330 K |
|---|---|---|---|---|---|
| P1 | 61.5±0.4 | | P18 | 237.7±0.4 | 230.7±1.7 |
| P2 | 69.5±0.6 | 61.1±1.2 | P19 | 247.3±0.3 | |
| P3 | 76.7±0.2 | 72.9±0.9 | P20 | 269.7±0.7 | |
| P4 | 83.3±0.09 | 82.7±0.5 | P21 | 290.0±0.5 | |
| P5 | 92.2±0.2 | 92.7±1.0 | P22 | 319.2±0.7 | |
| P6 | 96.5±0.1 | | P23 | 352.5±0.4 | 348.7±0.5 |
| P7 | 121.0±0.3 | | P24 | 364.4±0.3 | |
| P8 | 128.6±0.04 | 122.1±0.3 | P25 | 375.8±0.1 | |
| P9 | 137.6±0.1 | | P26 | 393.9±0.6 | 378.0±2.5 |
| P10 | 150.3±0.1 | | P27 | 402.4±0.1 | 394.9±0.8 |
| P11 | 157.5±0.1 | 150.3±0.4 | P28 | 429.7±0.5 | |
| P12 | 161.1±0.1 | | P29 | 454.7±1.0 | |
| P13 | 168.0±0.1 | | P30 | 481.9±0.9 | |
| P14 | 175.9±0.1 | 169.9±0.3 | P31 | 535.0±0.9 | 532.0±1.0 |
| P15 | 188.0±0.1 | 183.8±1.0 | P32 | 569.9±0.07 | 552.2±0.7 |
| P16 | 213.4±1.0 | | P33 | 643.9±1.4 | |
| P17 | 227.3±0.5 | | | | |

**Table S3:** List of the fitting parameters corresponding to the phonon modes in La$_4$Ni$_3$O$_{10}$, fitted using the three-phonon fitting model as described in the text. Units are in cm$^{-1}$.

| Modes | $\omega_0$ | A | $\Gamma_0$ | B |
|---|---|---|---|---|
| P8 | 130.5 ± 0.3 | -1.2 ± 0.1 | - | - |
| P11 | 162.3± 0.01 | -2.1± 0.01 | 3.2± 1.4 | 1.6± 0.3 |
| P14 | 178.4± 0.4 | -1.7± 0.1 | 5.7± 1.3 | 0.7± 0.4 |
| P18 | 246.4± 1.2 | -4.8± 0.5 | 8.9±0.01 | 4.7±0.3 |
| P23 | 356.5± 0.4 | -3.1± 0.2 | 4.2 ± 1.9 | 3.3 ± 1.0 |
| P26 | 409.1 ± 2.5 | -12.3± 1.6 | - | - |
| P27 | 408.8± 0.8 | -6.2± 0.5 | - | - |
| P32 | 590.8± 0.9 | -21.4± 0.8 | 2.0± 1.8 | 14.6± 1.5 |

**Table S4:** $\Gamma_n/\omega_n^2$ $(eV)^{-1}$ and $\lambda$ for the observed phonon modes at T=300 K and 100 K.

| Mode | T=300 K | | T=100 K | | $\Delta\lambda_{relative\ change}$ (%) |
|---|---|---|---|---|---|
| | $\Gamma/\omega^2$ $(eV)^{-1}$ | $\lambda$ | $\Gamma/\omega^2$ $(eV)^{-1}$ | $\lambda$ | |
| P8 | 3.3 | 0.55 | 1.21 | 0.19 | 65.5 |
| P11 | 3.6 | 0.59 | 1.43 | 0.222 | 62.4 |
| P14 | 2.9 | 0.41 | 1.31 | 0.16 | 60.9 |
| P15 | 2.39 | 0.391 | 1.04 | 0.16 | 59.1 |
| P18 | 4.09 | 0.671 | 1.143 | 0.177 | 73.6 |
| P23 | 0.92 | 0.15 | 0.51 | 0.079 | 47.3 |
| P27 | 1.11 | 0.18 | 0.48 | 0.074 | 58.9 |
| P32 | 0.69 | 0.12 | 0.41 | 0.061 | 49.2 |